# Optically Thick [O I] and [C II] Emission Towards NGC 6334 A

N. P. Abel[1], A. P. Sarma[2], T. H. Troland[3], & G. J. Ferland[3]

[1]Department of Physics, University of Cincinnati, Cincinnati, OH, 45221
npabel2@gmail.com

[2]Depaul University, Department of Physics, Chicago, IL, 60614,
asarma@depaul.edu

[3]University of Kentucky, Department of Physics and Astronomy, Lexington, KY 40506 troland@pa.uky.edu, gary@pa.uky.edu

## Abstract

This work focuses on [O I] and [C II] emission towards NGC 6334 A, an embedded $H^+$ region/PDR only observable at infrared or longer wavelengths. A geometry where nearly all the emission escapes out the side of the cloud facing the stars, such as Orion, is not applicable to this region. Instead, we find the geometry to be one where the $H^+$ region and associated PDR is embedded in the molecular cloud. Constant density PDR calculations are presented which predict line intensities as a function of $A_V$ (or $N$(H)), hydrogen density ($n_H$), and incident UV radiation field ($G_0$). We find that a single component model with $A_V$ ~650 mag, $n_H = 5\times10^5$ cm$^{-3}$, and $G_0 = 7\times10^4$ reproduces the observed [O I] and [C II] intensities, and that the low [O I] 63 to 146 micron ratio is due to line optical depth effects in the [O I] lines, produced by a large column density of atomic/molecular gas. We find that the effects of a density-law would increase our derived $A_V$, while the effects of an asymmetric geometry would decrease $A_V$, with the two effects largely canceling. We conclude that optically selected $H^+$ regions adjacent to PDRs, such as Orion, likely have a different viewing angle or geometry than similar regions detected through IR observations. Overall, the theoretical calculations presented in this work have utility for any PDR embedded in a molecular cloud.

## 1 Introduction

Photodissociation Regions, or Photon Dominated Regions (henceforth PDRs) mark the transition from ionized to atomic and molecular gas in star-forming regions. PDRs are usually physically adjacent to H II regions (which we refer to as $H^+$ regions in this work), although environments such as reflection nebulae (Hollenbach & Tielens 1997; Young Owl et al. 2002) and much of the diffuse ISM, which are exposed to few hydrogen-ionizing photons, also contain PDRs.

Overall, the bulk of mass in star-forming environments is contained in PDRs, and so these regions allow us to study important physical processes in astrophysical environments, such as grain physics, molecular formation, and the interplay between magnetic, thermal, turbulent, and gravitational energies (Hollenbach & Tielens 1997, Crutcher 1999).

In this paper, we focus our attention on a region where the assumed geometry used in most PDR calculations (such as Tielens & Hollenbach 1985 and Kaufman et al. 1999), fails to reproduce the [O I] emission line fluxes at 63, 146 μm and [C II] 158μm emission. In Section 2, we review the PDR diagnostic capabilities of [O I] and [C II] emission. Section 3 summarizes the observational data for NGC 6334 A. Section 4 gives the details of our theoretical calculations. Section 5 gives our results, and Section 6 gives our conclusions.

# 2 [O I] and [C II] Emission as a PDR Diagnostic

## 2.1 Background

Since their first detection ([O I] $^3P_1 \rightarrow {}^3P_2$; Melnick, Gull, & Harwit 1979; [O I] 146μm $^3P_0 \rightarrow {}^3P_1$; Stacey et al. 1983; [C II] 158μm $^2P_{3/2} \rightarrow {}^2P_{1/2}$; Russell et al. 1980 – all in Orion), emission from [O I] and [C II] has been widely regarded to come from PDRs. This is due to two reasons. One is that the ionization potentials of $O^0$ and $C^0$ mean that $O^0$ and $C^+$ are the dominant stages of ionization beyond the $H^+$ ionization front. The second reason is that the excitation energies for [O I] and [C II] emission range from 92 – 330 K, which are also the characteristic temperatures found in PDRs. Therefore, these transitions are easily excited by collisions with H and $H_2$.

Theoretical calculations have shown the diagnostic utility of [O I] and [C II] emission in many regions (Tielens & Hollenbach 1985). Such calculations involve combining current theories of atomic, molecular, and grain processes in PDRs with solving the problem of radiative transfer to determine what conditions reproduce the observed spectrum. PDR calculations typically have two free parameters, the total hydrogen density, $n_H$, and the strength of the UV radiation field between 6-13.6 eV ($G_0$, where 1$G_0$ equals $1.6\times10^{-3}$ erg cm$^{-2}$ s$^{-1}$, Habing 1968). Once an optimal $n_H$ and $G_0$ is found, the chemical and thermal structure of the PDR along with basic physical parameters such as mass, temperature, and energetics can be derived. Plane-parallel calculations like the ones presented in Kaufman et al. (1999) give contour plots showing how [O I], [C II] emission from the illuminated face (defined here as the side of the cloud facing the ionizing star or stars) along with other atomic/molecular emission line intensities vary over a wide range of $n_H$ and $G_0$. Such plots have been very useful in deriving physical conditions in a variety of galactic and extragalactic star-forming regions.

## 2.2 Limits to [O I] and [C II] Diagnostic Capability

Even with the success of PDR calculations in reproducing [O I] and [C II] observations, each line has unique features that diminish their diagnostic value. We describe the inherent problems with each line below:

***[O I] 63μm*** – Liseau, Justtanont, & Tielens (2006, henceforth LJT06) reviewed the results of the ISO mission and found that ~65% of the observed [O I] 63/146 emission line ratios (henceforth $[\mathrm{O\ I}]^{63\mu\mathrm{m}}_{146\mu\mathrm{m}}$) are lower than can be explained by current models (i.e. $[\mathrm{O\ I}]^{63\mu\mathrm{m}}_{146\mu\mathrm{m}} < 10$). LJT06 found that the likely cause of this discrepancy is optical depth effects in the 63μm line. LTJ06 partially attribute the low $[\mathrm{O\ I}]^{63\mu\mathrm{m}}_{146\mu\mathrm{m}}$ observations to absorption by cold, foreground $O^0$ in front of the 63μm emitting region, although other explanations such as masing in the 146μm line or very optically thick line emission cannot be ruled out. The overall conclusion from LJT06 is that, because the $[\mathrm{O\ I}]^{63\mu\mathrm{m}}_{146\mu\mathrm{m}}$ line depends sensitively on detailed models, [O I] emission has a limited diagnostic value.

***[O I] 146μm*** – Of the [O I] 63μm, 146μm and [C II] 158μm emission lines, the [O I] 146μm line is the hardest to detect in astrophysical environments. In a sample of 53 galaxies observed with ISO, Malhotra et al. (2001) detected the 146μm line in only 11 galaxies. This is four times fewer than the number of galaxies observed in 63μm emission. The faintness of this line diminishes its capabilities as a robust PDR diagnostic. The lower level of the 146μm line ($J = 1$) is not the ground state of $O^0$, meaning that [O I] 146μm emission is usually optically thin.

***[C II] 158μm*** – The [C II] line, more than either of the [O I] lines, can be emitted by $H^+$ regions. This is particularly true in low-density gas (Heiles 1994; Abel et al. 2005, Kaufman, Wolfire, & Hollenbach 2006; Abel 2006). This effect can hamper the use of [C II] emission as a pure PDR diagnostic in cases where ionized and PDR gas are observed in a single spectrum. Therefore, the contribution of [C II] emission from the ionized gas must be estimated. Such an estimate requires a separate model of the $H^+$ region, although in recent years computational methods exist that allow the $H^+$ region and PDR spectrum to be calculated self-consistently (Abel et al. 2005; Kaufman, Wolfire, & Hollenbach 2006). Therefore, even though [C II] emission is widely observed and is usually optically thin, its dependence on the properties of the $H^+$ region can diminish its use as a PDR diagnostic.

# 3 NGC 6334 A

The heavily obscured shell-like $H^+$ region NGC 6334 A is part of a star-forming complex 1.7kpc away. The observed [O I] and [C II] emission towards NGC 6334 A ($[\mathrm{O\ I}]^{63\mu\mathrm{m}}_{146\mu\mathrm{m}} = 2.4$ – see Table 1 and Kraemer, Jackson, & Lane 1998) cannot be explained with the geometry assumed in the widely used PDR calculations of Tielens & Hollenbach (1985) and Kaufman et al. (1999) (see Section 4.3). The close proximity and wealth of observational data for NGC 6334 A make it an excellent object to enhance our understanding of applying theoretical calculations to star-forming environments. In this section, we summarize our current understanding of this region.

## 3.1 $H^+$ Region

Conditions in the $H^+$ region that are derived from radio and IR observations are summarized in Table 1. Rodriguez et al. (1982) measured the flux at 6cm, which they found to be consistent with an O7.5 ZAMS star emitting $3\times10^{48}$ hydrogen ionizing photons per second ($Q_H$). The size of the $H^+$ region (see next paragraph) combined with the ionizing flux yields an electron density of $2\times10^4$ $cm^{-3}$.

Previous work in the radio and sub-millimeter has established dimensions for the $H^+$ region and the surrounding molecular cloud. Carral et al. (2002) mapped the region in the radio continuum at 3.5 cm. They interpreted their observations as a shell with a radius of 0.06 parsecs (15") and a thickness of 0.016 parsecs (~2").

## 3.2 PDR and Molecular Gas

Previous authors have observed many atomic and molecular line intensities, including [O I] (both 63μm and 146μm), [C II], and molecular transitions of CO, CS, OH, $NH_3$, and $H_2CO$. Observations of CO and CS in NGC 6334 A by Kraemer et al., 1997 were interpreted as a molecular torus with dimensions of $2.2\times0.9$ parsecs. In this interpretation, the molecular torus is elongated east-west, and surrounds the $H^+$ region. There are also magnetic field measurements (Sarma et al. 2000), due to the Zeeman splitting of OH at 18cm, which absorbs bremsstrahlung emission from the background $H^+$ region at frequencies of 1665 and 1667 MHz. The [O I] and [C II] observations from Kraemer, Jackson, & Lane (1998) are given in Table 1, with all three intensities scaled to the same beam size. Despite the breadth of observations, the physical conditions in the PDR are quite uncertain. Kraemer et al. (2000) and Sarma et al. (2000) find a total hydrogen column density $N(H)=N(H^0) + 2\times N(H_2) \sim 10^{23}$ $cm^{-2}$, although more recent data suggests it is probable that $N(H_{tot})$ exceeds $10^{24}$ $cm^{-2}$ (see 3.3) Both $n_H$ and $G_0$ are uncertain by over an order of magnitude. The primary cause of the uncertainty is that $[O\ I]^{63\mu m}_{146\mu m} = 2.4 \pm 0.5$. Kraemer, Jackson, & Lane (1998) note that an adequate PDR model for NGC 6334 A needs to consider the radiative transfer effects of [O I] and [C II] emission having to travel through the molecular cloud in order to reach our detectors (see Section 4.1 for a discussion of geometry considerations). Kraemer et al. (2000) therefore use CO and FIR observations, which are less sensitive to extinction, to infer physical conditions.

## 3.3 $A_V$ towards NGC 6334 A

It is clear that $A_V$ towards NGC 6334 A is very large, with the actual value being of some debate. If we assume a dust-to-gas ratio $A_V/N(H_{tot})$ = $5\times10^{-22}$ mag cm$^2$ (Savage et al. 1977), then $N$(H) = $10^{23}$ cm$^{-2}$ yields an $A_V$ ~ 50 mag. Harvey, Hyland, & Straw (1987) argue from near infrared observations that $A_V$ must be at least 50 mag towards the ionizing source, while Persi, Tapia, & Roth (2000) state that $A_V$ > 100 mag.

Recent 850μm flux maps from Sandell (1999) suggests a much higher $A_V$. Assuming the 850μm emission is optically thin, the optical depth at 850μm, $\tau_{850}$, equals the observed intensity divided by the Planck function $B_\nu = \frac{2h\nu^3}{c^2}\left(\frac{1}{\exp(h\nu/kT_d)-1}\right)$, where $T_d$ is the dust temperature. Since $\frac{h\nu}{kT_d} << 1$ at 850μm, $B_\nu \approx \frac{2k\nu^2 T_d}{c^2}$. Using the observed peak flux towards NGC 6334 A (15 Jy/beam, where the half power beams width, HPBW = 14.6"), we find:

$$\tau_{850} = I_{obs}/B_\nu = 8\times10^{-3}\left(\frac{100}{T_d}\right) \quad (1)$$

We can relate $\tau_{850}$ to $N(H_{tot})$ if we know how the dust extinction varies with λ. The dust opacity curve given in Draine & Lee (1984, Figure 9) for a mixture of silicate and graphite spherical grains yields $\tau_{850}$ = $9\times10^{-27}N(H_{tot})$. Equating the two expressions gives:

$$N(H_{tot}) = 9.0\times10^{23}\left(\frac{100}{T_d}\right) \quad (\mathrm{cm}^{-2}) \quad (2)$$

Assuming $A_V$ and $N(H_{tot})$ are related by the standard dust-to-gas ratio mentioned above, we get:

$$A_V = 450\left(\frac{100}{T_d}\right) \quad (\mathrm{mag}) \quad (3).$$

Equation 3 places useful constraints on the amount of extinction. Since dust sublimates around $10^3$ K, we can place a lower limit on $A_V$ of 45 mag. For typical $T_d$ = 50 – 150, $A_V$ range from 300 – 900 mag. This suggests that $N(H_{tot})$ may be an order of magnitude higher than previously thought.

# 4 Calculations and Results

Our calculations use the developmental version of the spectral synthesis code Cloudy, last described by Ferland et al. (1998), to determine whether radiative transport effects of the molecular cloud on the PDR spectrum can explain the [O I] and [C II] emission line ratios observed towards NGC 6334 A. We therefore do not consider molecular emission features, such as CO or OH.

## 4.1 Abundances, and Stopping Criteria

Our assumed gas-phase abundances are an average from the work of Cowie & Songalia (1986) and Savage & Sembach (1996). The two most important abundances for this calculation, oxygen and carbon, have assumed gas-phase abundances (relative to hydrogen) of $3.2\times10^{-4}$ and $2.5\times10^{-4}$, respectively. We assume a galactic ratio of visual extinction to hydrogen column density, $A_V/N(\mathrm{H_{tot}})$, of $5\times10^{-22}$ mag cm$^2$. Grain size distributions for gas adjacent to $H^+$ regions, such as Orion (Cardelli et al. 1989) tend to have a larger ratio of total to selective extinction than observed in the ISM. We therefore use a truncated MRN grain size distribution (Mathis et al. 1977) with $R$ = 5.5, which reproduces the Orion extinction curve (Baldwin et al. 1991). We also include size-resolved PAHs in our calculations, with the same size distribution used by Bakes & Tielens (1994). The abundance of carbon atoms in PAHs that we use, $n_C(\mathrm{PAH})/n_H$, is $3\times10^{-6}$. PAHs are thought to be destroyed by hydrogen ionizing radiation and coagulate in molecular environments (see, for instance, Omont 1986). We model this effect by scaling the PAH abundance by the ratio of $H^0/H_{tot}$ ($n_C(\mathrm{PAH})/n_H = 3\times10^{-6}\times [n(H^0)/n(H_{tot})]$).

We predict the PDR emission line spectrum for increasing hydrogen column densities, which we give in terms of $A_V$ (by assuming the gas-to-dust ratio given in 3.3). We initially stop the model when $A_V$ = 1 mag, and then increase $A_V$ by a factor of 1.5 until we reach an $A_V$ ~1500 mag. This allows us to consider the range of possible extinctions given by equation 3. Our calculations include a turbulent linewidth $\Delta v_{turb}$ = 3 km s$^{-1}$, which is consistent with the observed OH absorption linewidth towards NGC 6334 A observed by Sarma et al. (2000). The effect of $\Delta v_{turb}$ is to reduce the predicted line optical depth, since $\tau$ and linewidth are inversely proportional.

## 4.2 Radiation Field and Density

The shape and intensity of the UV continuum for NGC 6334 A is highly uncertain (see references in Section 3.1 and 3.2). Due to the uncertainty in the incident continuum, and because we are primarily interested in the transport of the PDR lines through the molecular cloud and not the exact nature of the continuum shape, we chose the widely used Draine (1978) UV radiation field in our calculations. The Draine continuum contains no hydrogen ionizing radiation, and is defined over the energy range 5 – 13.6eV. The relationship between the Draine field (henceforth $\chi$) and $G_0$ is $1G_0 = 1.71\chi$ (Draine & Bertoldi 1996). We consider two values for $G_0$, $10^{4.5}$ and $10^{5.5}$. These values are consistent with the range of $G_0$ given in Kraemer et al. (2000) of $10^{4.0}$ – $10^{6.2}$, based on the estimated $G_0$ from an O7.5 star and the PDR models of Kaufman et al. (1999).

To account for hot dust emission from the $H^+$ region, we include a $T$ = 75 K blackbody with a total intensity of $5\times10^2$ erg cm$^{-2}$ s$^{-1}$. This is identical to the dust continuum used by Tielens & Hollenbach (1985), used to model the dust emission coming from the Orion $H^+$ region. We found that, though the dust continuum increased the total luminosity of the system, it made little or no difference in the predicted [O I] and [C II] emission line intensities. The only other ionization source considered is cosmic rays. We include primary and secondary cosmic ray ionizations as described in Appendix C of Abel et al. (2005), with a cosmic ray ionization rate of $5\times10^{-17}$ s$^{-1}$.

Since our calculations do not consider hydrogen-ionizing radiation, we had to estimate the portion of [C II] emission which is due to the ionized gas. The high $n_e$ of the shell $H^+$ region, combined with the measured $Q$(H) and radius, suggests most of the [O I] and [C II] emission comes from the PDR and not the $H^+$ region (Abel et al. 2005, Kaufman, Wolfire, & Hollenbach 2006; Abel 2006, see also Table 1). We use the results of Abel (2006), which calculate the percentage of [C II] emission from $H^+$ regions as a function of $H^+$ density, ionization parameter $U$ ($U = \frac{Q_H}{4\pi r^2 n_e c}$), stellar temperature $T_*$, and stellar atmosphere. With the values of $Q_H$, $r$, and $n_e$ given in Table 1, we find $U \approx 10^{-1.9}$. Additionally, the observed $Q_H$ corresponds approximately to $T^*$ = 35,000 K (Vacca et al. 1996). Using these values and Figure 2 of Abel (2006), we find 5-15% of the total [C II] emission comes from the $H^+$ region. We therefore adopt the value of 90% for observed [C II] emission that comes from the PDR.

The remaining free parameter in our calculations is $n_H$. Our primary interest is to determine if a single component model for NGC 6334 A can explain the observed [O I] and [C II] emission. We therefore assume that $n_H$ is constant throughout our calculations, and consider values of $10^{4.0}$, $10^{5.0}$, $10^{6.0}$ cm$^{-3}$. Such a density profile is an approximation to regions where magnetic or turbulent pressure dominates (Tielens & Hollenbach 1985). A real molecular cloud is not homogeneous, but rather a clumpy, dynamically evolving entity. Our choice of $n_H$ therefore represents an average over the physical extent of the cloud. We explore the effects of a density power-law on the best-fitting single component calculation in Section 5.2. Overall, our choice of $n_H$, $G_0$, and stopping $A_V$ represents a total of 114 different calculations.

The most detailed calculation for NGC 6334 A would also include the ionized gas, and connect the $H^+$ region to the PDR through an equation of state such as constant pressure (Abel et al. 2005; Kaufman, Wolfire, & Hollenbach 2006). There are two reasons we chose to ignore the $H^+$ region in this work. For one, we wanted to compare our results to other PDR calculations, in particular Tielens & Hollenbach (1985), Kaufman et al. (1999), and LJT06, all of which do not include the $H^+$ region. Secondly, we wanted to demonstrate an application of Abel et al. (2005) and Abel et al. (2006) in estimating the % emission of [C II] from the $H^+$ region, if both $U$ and $n_e$ are known. The proximity and amount of observational data for NGC 6334 A makes the region an excellent case study.

## 4.3 Geometry Considerations

The usual geometry assumed in PDR calculations is a single-sided plane-parallel slab (see Tielens & Hollenbach 1985a; geometry shown in Figure 1A). We will refer to this geometry as an "open geometry". An open geometry is a good approximation to "blister $H^+$ regions" adjacent to PDRs (Tielens & Hollenbach 1985b) such as Orion. For an open geometry, it is convenient to define the direction towards the star as "inward" and the direction away from the star as "outward" (see Figure 1A). Using this nomenclature, the [O I] 63 μm emission is very optically thick in the outward direction, forcing all the [O I] emission to become beamed in the inward direction. The [O I] 145 μm and [C II] 158 μm lines have smaller optical depths, therefore allowing a small fraction of photons to escape in the outward direction. The predicted line intensity is presented as the emergent flux in the inward direction, divided by $2\pi$ (Kaufman et al. 1999).

The other possible geometry is one where the PDR emitting region is completely surrounded by a large column density of molecular gas. We will refer to this geometry as a "closed geometry". A similar geometry is used in Doty & Neufeld (1997) in the modeling of dense molecular cores. In a closed geometry, all energy eventually escapes in the direction outward from the star.

For both the open and closed geometries, we assume the velocity field of the PDR is the same as the molecular cloud. We refer to this assumption as a "static geometry". Assuming a static geometry is justified since the observed linewidths for H I and OH absorption, along with $H_2CO$ emission, are nearly identical (De Pree et al 1995; Sarma et al. 2000). This indicates the broadening of both regions is dominated by turbulence.

It is improbable that the PDR emitting region is exactly at the center of the molecular cloud. Instead, the true geometry is likely to be asymmetric, with the ionizing stars being closer to one side of the molecular cloud (Figure 1C). Observations suggest that the $H^+$ region is located closer to the back side than to the front side of the molecular cloud. In particular, the radio frequency recombination line observations of DePree et al. (1995) show that the southern lobe of ionized gas is redshifted relative to the central $H^+$ region by about 10 km $s^{-1}$. Such a redshift is most naturally explained if the $H^+$ region is located near the back side of the cloud and has opened a hole in the cloud through which ionized gas is expanding away from the observer. We discuss the implications of an asymmetric geometry where the thicker part of the molecular cloud is along our line of sight in Section 5.2.

## 4.4 Open Geometry Results

It is possible that NGC 6334 A is simply an open geometry, such as a blister $H^+$ region, where we observe the $H^+$ region through the molecular cloud. Such a scenario would be akin to observing Orion rotated 180 degrees, from the perspective of being on the opposite side of OMC-1. If true, then this geometry must lead to an outward $[\mathrm{O\ I}]^{63\mu\mathrm{m}}_{146\mu\mathrm{m}}$ = 2-3 for physically plausible values of $A_V$, as defined by equation 3.

We performed an open geometry PDR calculation to determine if the outward $[\mathrm{O\ I}]^{63\mu\mathrm{m}}_{146\mu\mathrm{m}}$ ratio could be due to emission from a blister $H^+$ region on the far side of NGC 6334 A. The parameters for the model were $n_H = 10^5$ cm$^{-3}$ and $G_0 = 10^5$, with abundances and other physical parameters treated as described in Sections 4.1 and 4.2.

Figure 2 shows the predicted [O I] intensities in the inward and outward directions, along with the total emission. Figure 2 shows just how sensitive the 63 μm line is to optical depth effects. For extremely low values of $A_V$, some 63 μm emission escapes through the outward direction. For $A_V > 5$ however, the [O I] line becomes optically thick and over 90% of the 63 μm emission is beamed in the inward direction. The [O I] 145 μm remains essentially optically thin throughout, with the inward emission only slightly greater than the outward component.

The combined radiative transfer effects of the two [O I] lines are shown by Figure 3, which plots $[\mathrm{O\ I}]^{63\mu\mathrm{m}}_{146\mu\mathrm{m}}$ for the inward direction, outward direction, and total. The inward direction, corresponding to the geometry assumed in most PDR calculations, predicts a constant $[\mathrm{O\ I}]^{63\mu\mathrm{m}}_{146\mu\mathrm{m}}$. The predicted inward $[\mathrm{O\ I}]^{63\mu\mathrm{m}}_{146\mu\mathrm{m}}$ also exceeds the optically thin limit of 10. The outward $[\mathrm{O\ I}]^{63\mu\mathrm{m}}_{146\mu\mathrm{m}}$ ratio, which is what would be observed towards NGC 6334 A, is orders of magnitude lower, and depends sensitively on $A_V$.

Figure 3 essentially rules out an open geometry. If NGC 6334 A were a blister $H^+$ region observed through the molecular cloud, then the observed $[\mathrm{O\ I}]^{63\mu\mathrm{m}}_{146\mu\mathrm{m}}$ ratio would imply an $A_V$ of 10 – 20 mag, which is incompatible with the observed 850 μm emission (equation 3), as it would yield $T_d = (2\text{-}4)\times10^3$ K which exceeds the dust sublimation temperature. This ratio should depend primarily on the amount of extinction, and only weakly on $n_H$ or $G_0$. The extinction will depend somewhat on line width ($\Delta v$) and the O/H abundance ratio. A turbulent linewidth of 3 km s$^{-1}$ was used in the calculation, and is based on observations (Section 4.2). Therefore, the only free parameter that could alter this conclusion is the O/H ratio, which would have to be ~$10^{-6}$ in order to permit a high enough $A_V$ to be consistent with the 850 μm emission. Such an O/H ratio is unphysically low.

## 4.5 Closed Geometry Results

The results of our closed geometry calculations are shown in Figures 4-7 and Table 2. Table 2 gives the set of parameters ($n_H$, $G_0$, and $A_V$) that best-fits the observed spectrum, along with the predicted [O I] and [C II] emission-line intensities.

Figures 4-6 show the predicted [O I] and [C II] emission line intensities as a function of $n_H$, $G_0$, and $A_V$. Increasing either $n_H$ or $G_0$ increases the line intensity, which is typical for PDR calculations (Tielens & Hollenbach 1985; Kaufman et al. 1999). As the density is decreased, the size of the [O I] emitting region increases. This is because the size of the $H^0$ region, which is where [O I] and [C II] emission forms, is proportional to the ratio $G_0/n_H$. Lowering $n_H$ results in less $H_2$ shielding, producing a larger $H^0$ region and shifting the peak emission to larger depths (typically $A_V$ = 2-10, with 10 corresponding to Log[$n_H$] = 4). For larger $A_V$, line optical depths reduce the emergent intensity. As expected, line optical depth effects are largest for the [O I] 63μm emission line (Figure 2). Our calculations show the 63μm intensity decreases by over two orders of magnitude, with $\tau_{63\mu m}$ = 250 for $A_V$ = 1500 mag. The large optical depth is due the large $O^0$ ($J$=2) column density and not due to either self-absorption or absorption by dust.

This decrease in emergent intensity of the [O I] 63μm line is due to the large optical depth of the line as the $O^0$ column density increases. Increasing $\tau$ reduces the critical density ($n_{crit}$) of the 63μm line by $\tau$ (Osterbrock & Ferland 2006). For the 63μm line, $n_{crit} = 5\times10^5$ cm$^{-3}$, which means that, for the range of densities considered, $n_H > n_{crit}$ for large $A_V$. We refer to this effect as the line becoming "thermalized", as the large optical depth leads to the $^3P_1$ state preferentially de-exciting through collisions instead of through photon emission. In addition to thermalization of the 63μm line, there is also the fact that, for large $\tau$, the line will emit like a blackbody at the local temperature (typically a few tens of Kelvin). We can only see the 63μm line emerging from regions of a few optical depths. Therefore, for increasing $A_V$, the temperature decreases and the emergent intensity of the 63μm line is reduced.

Line optical depths are less important for the [O I] 146μm and[C II], although both lines do eventually become optically thick. At $A_V$ = 1500 mag, $\tau \sim 2.5$ for [O I] 146μm and $\tau \sim 1.6$ for [C II]. The reason for the lower optical depth is that there are much fewer $O^0$ atoms in the $J$=1 state of $O^0$ capable of increasing the 146μm optical depth. Extinction effects for [C II] are similar to [O I] 146μm, with one important difference. Carbon is primarily in the form of $C^+$ in $H^0$ regions, with $C^+$ converted into $C^0$ and CO beyond the $H^0/H_2$ molecular front. This leads to optical depth effects reducing 158μm emission in this region, which manifests itself as a decrease in the emergent intensity just beyond the peak. Once $C^+$ is converted to $C^0$ and CO, the effects of extinction are reduced, and only become significant when $A_V$ is large.

Figure 7 shows how $[O\ I]^{63\mu m}_{146\mu m}$ ratio varies with increasing $A_V$. The dependence of $[O\ I]^{63\mu m}_{146\mu m}$ on $G_0$ is weak, which is expected for the range of $n_H$ and $G_0$ we consider (see Figure 5 of Kaufman et al. 1999). For low $A_V$, $[O\ I]^{63\mu m}_{146\mu m}$ is a factor of three higher for the Log[$n_H$] = 4 case than for densities of Log[$n_H$] = 5 or 6. This is because, for Log[$n_H$] = 4, the [O I] line emitting region extends out $A_V \sim 10$. Therefore, to compare our calculations to Kaufman et al. (1999), we need to compare the predicted $[O\ I]^{63\mu m}_{146\mu m}$ ratio at the depth where both lines have fully formed, but have not yet suffered from extinction effects. This depth corresponds to the peak [O I] emission shown in Figures 4 & 5. As mentioned above, the peak [O I] emission for Log[$n_H$] = 4 is $A_V \sim 10$, while for Log[$n_H$] = 5 or 6 the peak is located around $A_V \sim$ 2-4. Using Figure 7, we find $[O\ I]^{63\mu m}_{146\mu m}$ is nearly constant for the depth where [O I] emission peaks. This also agrees fairly well with Figure 5 of Kaufman et al. (1999). Beyond the peak, the increased optical depth for [O I] 63μm (see Figure 4) causes $[O\ I]^{63\mu m}_{146\mu m}$ to fall below the optically thin limit of 10 (LJT06).

Our calculations allow us to find a set of physical parameters that reproduce the observed 1σ [O I] and [C II] emission-line spectrum (horizontal gray bars on Figures 4-7). Looking at Figure 7, we can see that we need $A_V$ to be between 400 - 800 mag in order to reproduce the observed $[O\ I]^{63\mu m}_{146\mu m}$. Figures 4 and 5 eliminates $n_H = 10^{4.0}$ cm$^{-3}$, since the predicted intensity is too low given the requirement for $A_V$. Figure 4-7 show that a $G_0 = 10^{4.5}$, $n_H = 10^{6.0}$ cm$^{-3}$, and $A_V \sim$ 600 - 700 mag model approximately reproduces the [O I] and [C II] emission. Using these values as a guide, we varied $n_H$, $G_0$, and $A_V$ around these values to find a set of $n_H$, $G_0$ that reproduces the observed intensities to within 2σ, with our derived quantities. The best values are $n_H = 10^{5.7}$ cm$^{-3}$, $G_0 = 10^{4.8}$, and $A_V$ = 650 mag, and are given in Table 2.

Figure 8 shows the gas temperature, [O I] 63μm optical depth, and the fraction of O and C in the form of $O^0$ and $C^+$ for the best fit parameters given in Table 2. The temperature decreases with increasing $A_V$, falling well below the excitation temperature for the two [O I] emission lines for $A_V$ > 2-5, meaning nearly all the emission occurs for $A_V$ < 5 mag. The fractional abundance of O in atomic form remains significant throughout the cloud complex, although the formation of CO reduces the $O^0$ abundance to 20% of its initial value. We included condensation of molecules onto grain surfaces (see Abel et al. 2005 for a description of how this is incorporated into Cloudy) in our modeling, which made no difference in the $O^0$ abundance with depth. The combination of low temperature and high $O^0$ abundance leads to a significant ground state $O^0$ column density, which in turn yields a large [O I] 63μm optical depth. The increase in optical depth with $A_V$ is not linear because the formation of CO, which decreases the $O^0$ abundance and therefore the ratio of [O I] 63μm optical depth per $A_V$.

We can compare our derived values of $n_H$ and $A_V$ with LTJ06 calculations of optically thick [O I] emission. Figure 4 of LTJ06 shows $[O\ I]^{63\mu m}_{146\mu m}$ as a function of temperature in the case where both [O I] lines are either optically thick or optically thin. The calculations presented in this work show the transition from optically thin emission, to optically thick [O I] 63μm emission - optically thin [O I] 146μm emission, and finally to the regime where both lines are optically thick. Figure 5 shows the value of $N(H^0)$ necessary to get an [O I] 146μm optical depth of unity. LTJ06 predicts the NGC 6334 A observed $[O\ I]^{63\mu m}_{146\mu m}$ ratio for a temperature of ~50K. At this temperature, LTJ06 predicts optically thick [O I] 146μm for $N(H^0) \sim 10^{24}$ cm$^{-2}$ and $n_H = 10^{55-6.0}$. Our results therefore appear to agree with the calculations shown in LTJ06.

Our results show that the observed [O I] and [C II] spectrum can be explained by a constant density PDR calculations combined with a closed geometry where the PDR emission must travel through a large column density of molecular gas. The derived $A_V$ = 650 mag corresponds to a dust temperature of 70 K. This $A_V$ also corresponds to $N(H_{tot}) = 10^{24.1}$ cm$^{-2}$, which corresponds to a physical thickness $L = N(H_{tot})/n_H$ of ~0.8 parsecs, within 20-30% of the size of the molecular torus in the plane of the sky (1.1 parsecs—see Figure 1). Overall, the density, radiation field, column density, and physical thickness are consistent with previous studies of the region.

# 5 Sensitivity to Model Assumptions

## 5.1 Effects of Density Law

While a single component PDR model can explain the observed [O I] and [C II] emission, it is highly unlikely that the density is uniform throughout the entire molecular complex. At the very least, the line of sight towards NGC 6334 A is clumpy (Kraemer et al. 1997). Both observational and theoretical evidence suggests the density in molecular clouds follow some power law dependence on density ($n \propto r^{\alpha}$, see Doty & Neufeld 1997 and references therein), with $\alpha \sim -2$. In the study of 14 envelopes around massive stars, van der Tak et al. (2000a) found $\alpha \sim -1.0, -1.5$ best fit the observations. A density gradient will change the predicted chemical/thermal structure of the cloud, which will have consequences for the predicted [O I] and [C II] emission.

We explored the effects of a density law on the best-fitting, constant-density model. We chose a density law of the form:

$$n(H)_r = n(H)_0 \left(1+\frac{r}{r_c}\right)^{\alpha} \text{ cm}^{-3} \qquad (4)$$

where $r$ is the depth into the cloud, $r_c$ is the thickness over which the density remains relatively constant, and $n(\mathrm{H})_0$, $n(\mathrm{H})_r$ are the hydrogen densities at the illuminated face and at a given depth, respectively. Density laws of this form are commonly assumed in the study of hot molecular cores (see Nomura & Millar 2004 and references therein).

We constrained our density law parameters in order to make a direct comparison between the best-fitting constant density models. We considered four values for $\alpha$; -2, -1.5, -1, and -0.5. We required that each calculation reproduce $N(\mathrm{H})$ (and hence $A_V$) given by the best single component model. We also required that the model stop at the thickness given by the single component model, which allows us to compare calculations with the same average density as the single component model. We chose $n(\mathrm{H})_0 = 10^6$ cm$^{-3}$, which would be the PDR surface density if the $H^+$ region and PDR were in gas pressure equilibrium with $T(H^+) = 10^4$ K and $T$(PDR) = 200 K. There is a unique value of $r_c$ (for each $\alpha$) which satisfies our constraints, and these values are given in Table 3.

The results of our density law calculations are given in Table 3. We find that a density law increases the predicted PDR line intensities, with a 0.11 dex increase in $I_{[\mathrm{O\,I}]\,63\mu\mathrm{m}}$, a 0.06 dex increase in $I_{[\mathrm{O\,I}]\,145\mu\mathrm{m}}$, and a 0.02 dex increase in $I_{[\mathrm{C\,II}]\,158\mu\mathrm{m}}$. However, we also find, for $\alpha \neq 0$, that the predicted intensities are independent of $\alpha$. This is due to the fact that, for all $\alpha$, the [O I] and [C II] emission emerges from a region with the same value of $n_H$ ($10^6$ cm$^{-3}$) and $G_0$ ($10^{5.2}$). Additionally, since we kept the total column density fixed, the effects of extinction were nearly identical for each model. If we had chosen the density from the constant density model as $n(\mathrm{H})_0$ and stopped at $A_V$ = 650 mag, we would have reproduced the observed PDR spectrum. However, the physical thickness required to reproduce an $A_V$ this high is much larger than the constant density model (10 parsecs for $\alpha$ = -2 and 3 parsecs for $\alpha$ = -0.5; compared to 0.8 parsecs for the constant density model). Such a model would also yield a lower average $n(\mathrm{H})$, $\overline{n(H)}$, although the average would still exceed $10^5$ cm$^{-3}$.

We performed a final test to see if a density-law model could reproduce the observed PDR emission line spectrum. We took the $\alpha$ = -2 power law with $n(\mathrm{H})_0 = 10^6$ cm$^{-3}$ and allowed $A_V$ to increase beyond the single component model $A_V$ = 650 mag. Table 4 summarizes our results. Increasing $A_V$ increases the line optical depths, thereby reducing the emergent intensity. For $A_V$ = 700 – 750, the density law and constant density model predict nearly identical [O I] and [C II] line intensities, which in turn agree with observations. Therefore, we conclude that the effects of a density law are to increase $A_V$ (or $N(\mathrm{H})$).

## 5.2 Effects of an Asymmetric Geometry

The closed and open geometry are the two limiting geometries, while an asymmetric geometry (Figure 1C) is the intermediate case. Section 4.4 showed that, in a pure open geometry, an $A_V$ of 20 between us and the ionizing stars reproduces the observed $[O\ I]^{63\mu m}_{146\mu m}$, while $A_V$ = 650 reproduces the observed $[O\ I]^{63\mu m}_{146\mu m}$ for a closed geometry. As argued in Section 4.3, the asymmetric case is one where more molecular material lies between us and the ionizing stars (near side) than between the ionizing stars and the back of the molecular cloud (back side). The asymmetric case differs from the open geometry in that the back side of the cloud no longer has radiation freely escaping. Instead the back side can also become optically thick to [O I] 63 μm emission. This leads to more [O I] emission escaping out the near side, increasing the observed $[O\ I]^{63\mu m}_{146\mu m}$. Therefore, the amount of $A_V$ needed to reproduce the observed $[O\ I]^{63\mu m}_{146\mu m}$ for the asymmetric geometry is greater than the pure open case ($A_V > 20$). The asymmetric geometry also differs from the closed geometry in an important way. In a closed geometry, the [O I] 63 μm optical depth towards the near or back side is the same. In the asymmetrical case, the optical depth in the back side of the cloud is less than in the near side. This leads to more [O I] 63 μm emission escaping out the back side than the near side, which lowers $[O\ I]^{63\mu m}_{146\mu m}$. Therefore, the amount of $A_V$ needed to reproduce the observed $[O\ I]^{63\mu m}_{146\mu m}$ for the asymmetric geometry is less than the pure closed case ($A_V < 650$).

Overall, our results place several constraints on $A_V$. Based on geometrical considerations, $A_V$ can range from 20 – 650, with 650 being the best-fit model to the closed geometry, constant density scenario. If we make the assumption that $T_d < 150$ K, then equation 3 constrains $A_V$ to values of 300 – 650. Taking the effects of a power law density structure into account, then the observed $[O\ I]^{63\mu m}_{146\mu m}$ can be reproduced with $A_V$ as large as 750. However, if we consider an asymmetric geometry, then the observed $[O\ I]^{63\mu m}_{146\mu m}$ can be modeled with $A_V < 650$. Therefore, the systematic effects of density structure and asymmetry largely neutralize each other. Given these facts, we feel that our conclusions about the physical properties of the NGC 6334 A PDR are fairly robust.

# 6 Conclusions

In this paper, we have presented a series of constant density, PDR models of NGC 6334 A. Overall, we find the following:

- We considered both single-sided, plane-parallel (i.e. open geometry, such as Orion) and one where the PDR emitting region is completely surrounded by molecular gas (closed geometry). We find the Orion geometry, where the PDR emission is beamed away from us, yields an $A_V$ which is incompatible with the 850μm emission in the region.

- We find that a closed geometry is able to simultaneously reproduce the observed [O I] and [C II] emission with realistic values of $A_V$. Optical depth effects decrease the $[O\ I]^{63\mu m}_{146\mu m}$ ratio with increasing $A_V$. Our best-fitting model yields $A_V = 650$ mag, $G_0 = 10^{4.8}$, and $n_H = 10^{5.7}$ cm$^{-3}$ giving the best agreement with observation. Additionally, our results agree well with the results of LTJ06, in the limit where both [O I] lines become optically thick.
- Optically selected objects, like the Orion H II region, tend to be viewed face on, the geometry assumed in most PDR calculations. But IR selected objects may be viewed from any angle, and may be embedded within molecular gas. Therefore, modeling IR selected regions may often require a different geometry than assumed by open geometry models such as Kaufman et al. (1999).
- We show the effects of a density law on the best-fitting model are to increase $A_V$ to ~ 750 mag, or to increase the cloud thickness. If the geometry is asymmetric rather than symmetric, $A_V$ is < 650. Overall, these two effects largely negate each other.
- Overall, the calculations outlined here should prove useful in studying heavily embedded star-formation, which will be observed by Spitzer, Herschel, SOFIA, and ALMA with regularity.

Acknowledgements: NPA would like to acknowledge Center for Computational Sciences at the University of Kentucky for financial and computational support, along with NSF grant #0094050. NPA and APS acknowledge support from Spitzer GO-2 program #20220. GJF acknowledges NSF through AST 06-07028 and NASA through NNG05GD81G, THT acknowledges support for this project from NSF grant AST 03-07642. We thank the anonymous referee for a careful reading on this manuscript and her/his many useful comments.

# 8 Tables

Table 1 Observational Data

| Observable | Value | Reference |
|---|---|---|
| Log[$I_{[O\ I]\ 63\mu m}$] (erg cm$^{-2}$ s$^{-1}$) | -1.66 (-1.65 to -1.70) | Kraemer, Jackson, & Lane (1998) |
| Log[$I_{[O\ I]\ 146\mu m}$] (erg cm$^{-2}$ s$^{-1}$) | -2.04 (-1.97 to -2.12) | Kraemer, Jackson, & Lane (1998) |
| [1]Log[$I_{[C\ II]\ 158\mu m}$] (erg cm$^{-2}$ s$^{-1}$) | -1.96 (-1.93 to -2.01) | Kraemer, Jackson, & Lane (1998) |
| $Q_H$ (photons s$^{-1}$) | $3\times10^{48}$ | Rodriguez et al. (1982) |
| $n_e$ (cm$^{-3}$) (H$^+$ region) | $2\times10^4$ | Carral et al. (2002) |
| Size of H$^+$ region (cm) | $1.9\times10^{17}$ | Carral et al. (2002) |
| [2]$A_V$ (mag) | $445\left(\frac{100}{T_d}\right)$ | Sandell (1999) and this work |

[1]A 10% correction was applied for H$^+$ region component to [C II] emission

[2]See equation 3

Table 2 Best-Fit Constant Density Model

| Quantity | Value |
|---|---|
| Log[$I_{[O\ I]\ 63\mu m}$] | -1.67 |
| Log[$I_{[O\ I]\ 146\mu m}$] | -2.02 |
| Log[$I_{[C\ II]\ 158\mu m}$] | -1.95 |
| $n_H$ | $10^{5.7}$ |
| $G_0$ ($1G_0 = 1.71\chi$) | $7\times10^4$ |
| $N$(H) | $1.3\times10^{24}$ (cm$^{-2}$) |
| Physical Thickness ($L$) | 0.8 parsecs |
| $A_V$ | 650 mag |

Table 3 Dependence of NGC 6334 A Model on Density Law

| Quantity | $\alpha$ = -2 | $\alpha$ = -1.5 | $\alpha$ = -1 | $\alpha$ = -0.5 | $\alpha$ = 0 |
|---|---|---|---|---|---|
| [1]$r_c$ (pc) | 0.95 | 0.65 | 0.30 | 0.10 | n/a |
| Log[$I_{[O\ I]\ 63\mu m}$] | -1.56 | -1.56 | -1.56 | -1.56 | -1.67 |
| Log[$I_{[O\ I]\ 146\mu m}$] | -1.96 | -1.96 | -1.96 | -1.96 | -2.02 |
| Log[$I_{[C\ II]\ 158\mu m}$] | -1.93 | -1.93 | -1.93 | -1.93 | -1.95 |

[1]Density law given by equation 4

Table 4 Dependence of $\alpha$ = -2 NGC 6334 A Model on Extinction

| $A_V$ (mag) | Log[$I_{[O\ I]\ 63\mu m}$] | Log[$I_{[O\ I]\ 146\mu m}$] | Log[$I_{[C\ II]\ 158\mu m}$] |
|---|---|---|---|
| 550 | -1.41 | -1.92 | -1.90 |
| 600 | -1.49 | -1.94 | -1.92 |
| 650 | -1.56 | -1.96 | -1.93 |
| 700 | -1.63 | -1.98 | -1.95 |
| 750 | -1.70 | -1.99 | -1.97 |

# 9 Figures

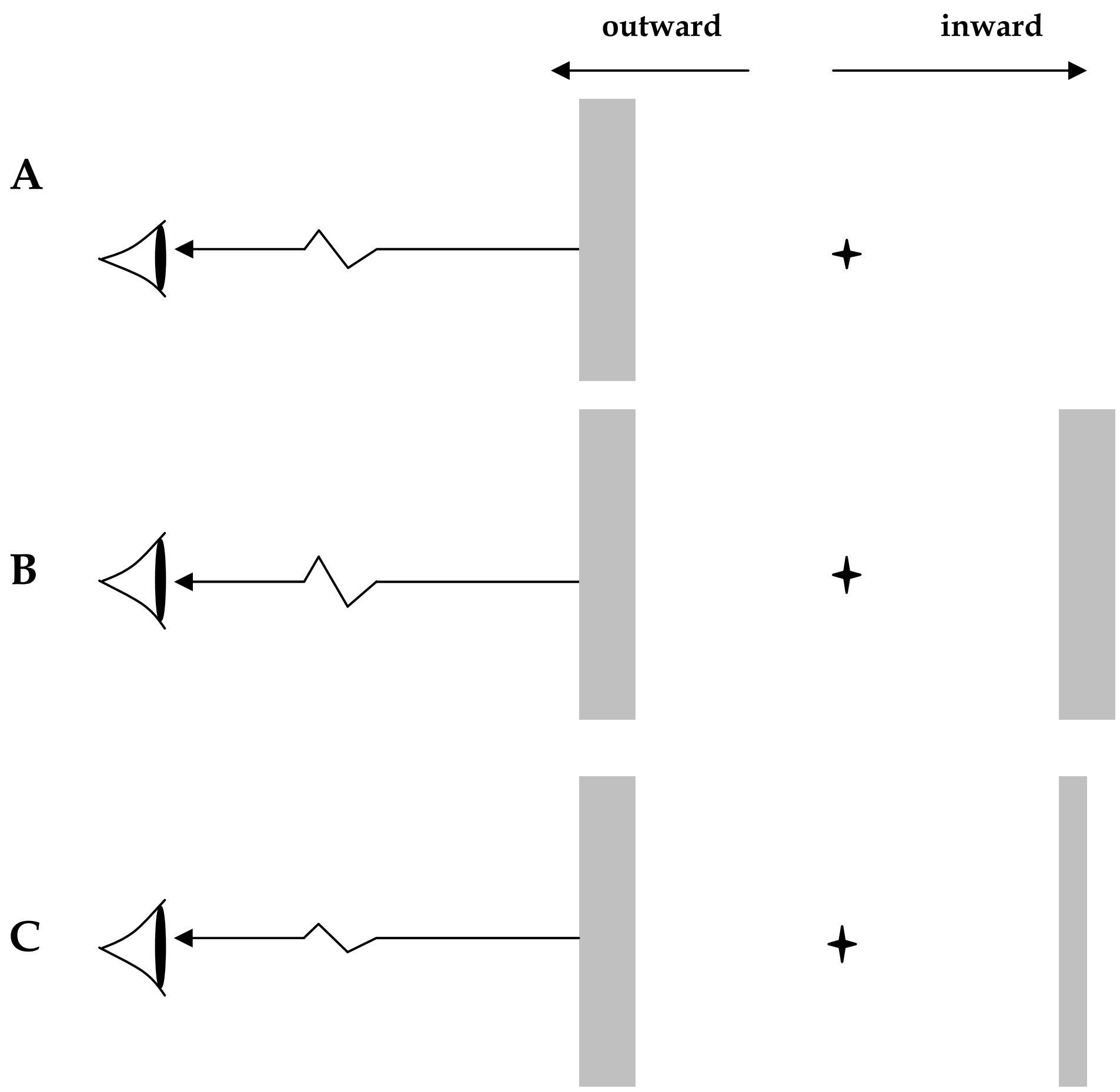

Figure 1 The three possible geometries for NGC 6334 A considered in this paper. These scenarios are: an open geometry where radiation freely escaped out the hot, illuminated face (inward direction), but is optically thick in the outward direction (A), a closed geometry where the radiation is smothered on both sides by a thick molecular cloud (B), or an asymmetric geometry where the molecular cloud is not as thick on one side (C). Our calculations presented in Section 5 assume geometry B.

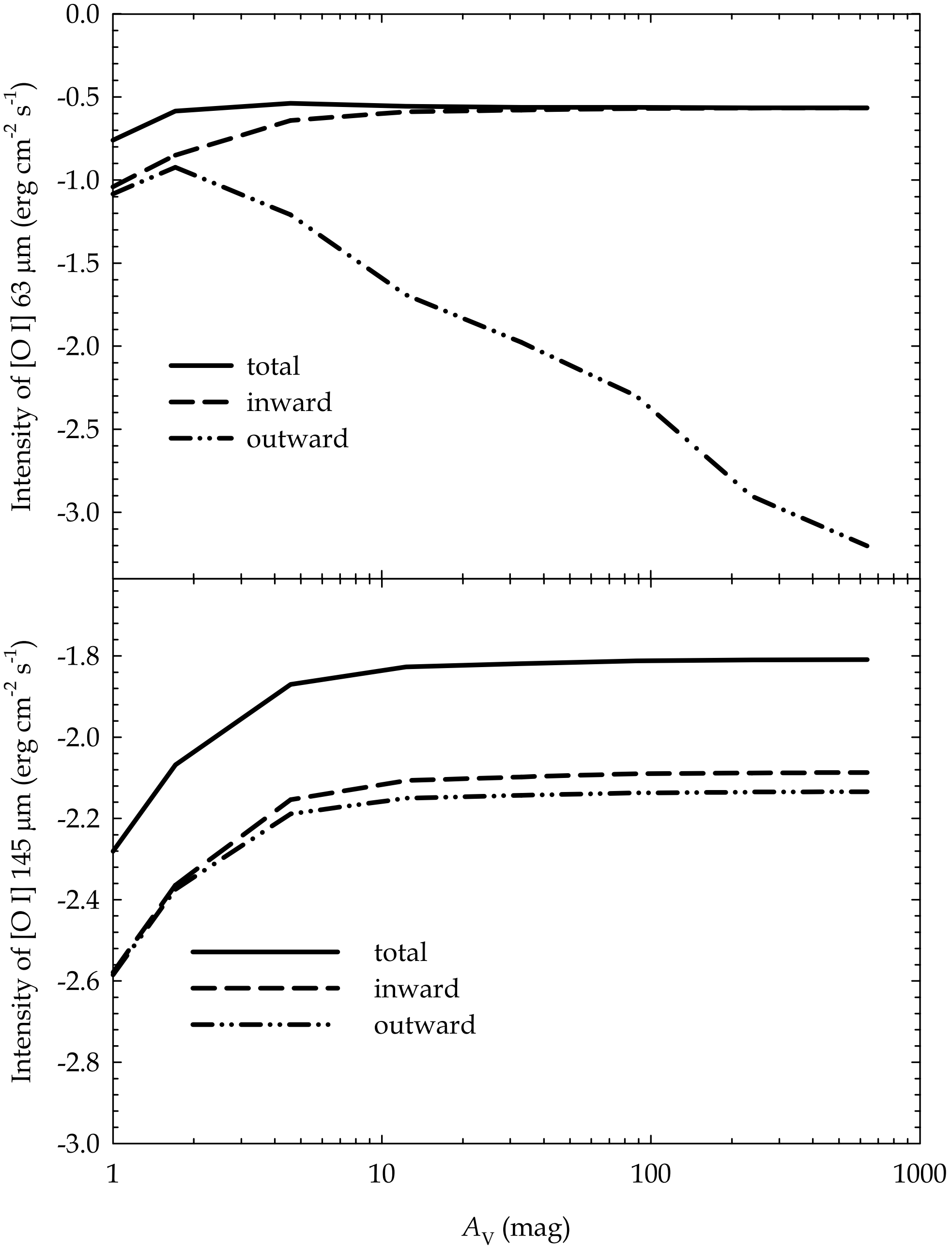

Figure 2 Intensity of [O I] lines as a function of $A_V$, for the open geometry shown in Figure 1A. The outward component of the 63 μm line falls off rapidly with $A_V$, due to the large optical depth in the outward direction, which beams all 63 μm emission in the inward direction. The 145 μm line has a much smaller optical depth in the outward direction, meaning it can escape freely in either direction. This is shown by the inward and outward intensities being approximately equal.

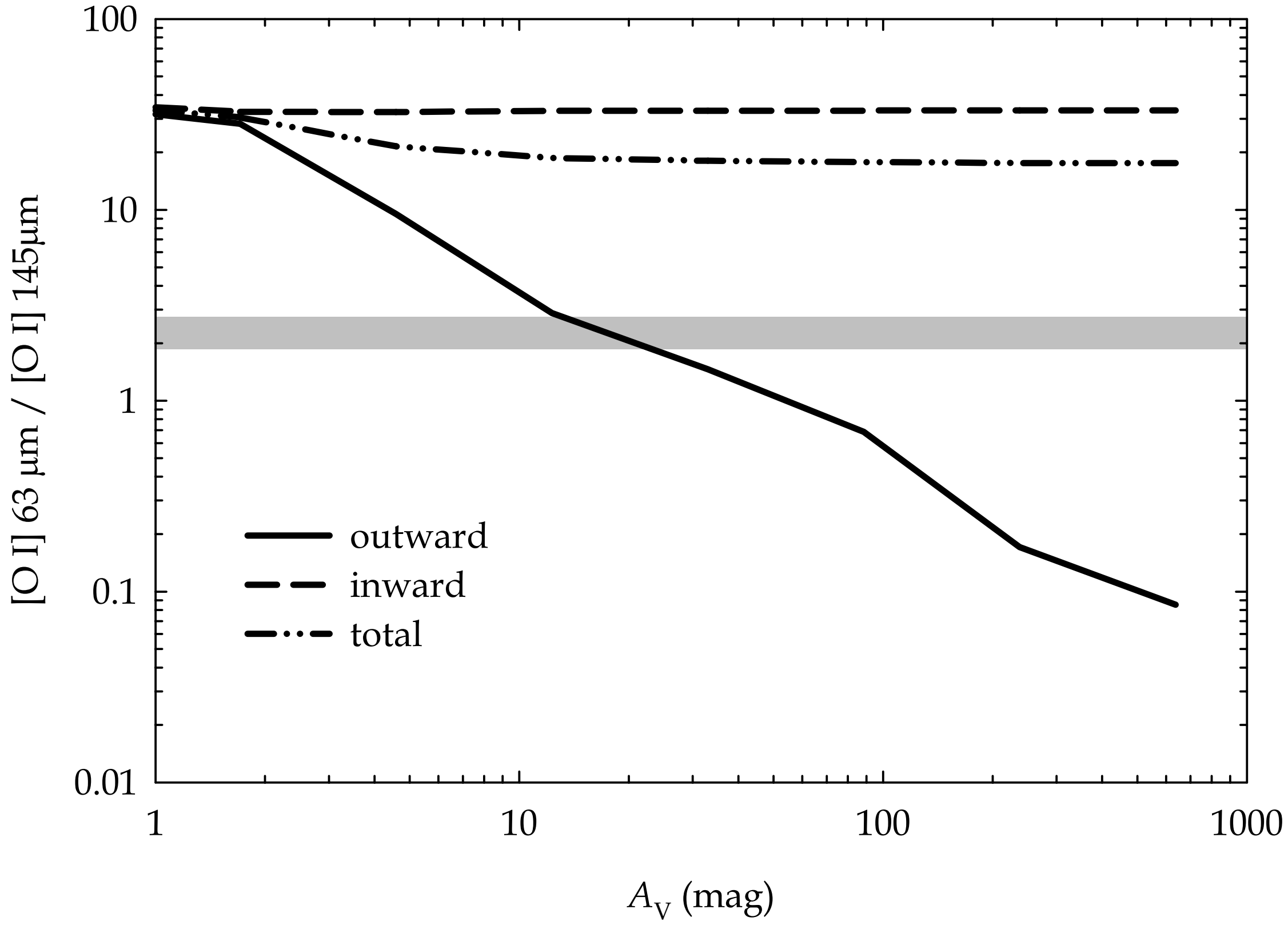

Figure 3 Ratio of [O I] 63/146 intensities in the inward and outward directions, along with the total, for the open geometry shown in Figure 1A. The inward directions shows the ratio in the optically thin inward direction remaining >10, which is what is usually reported in plane-parallel PDR calculations. Since all 63 μm emission is beamed in the outward direction (Figure 2), 63/146 decreases dramatically with $A_V$. The horizontal gray bar represents the observed ratio towards NGC 6334 A. This shows that for $A_V$ > 30 mag, the predicted ratio falls below the observed limits. This low $A_V$ is incompatible with 850 μm observations (equation 3).

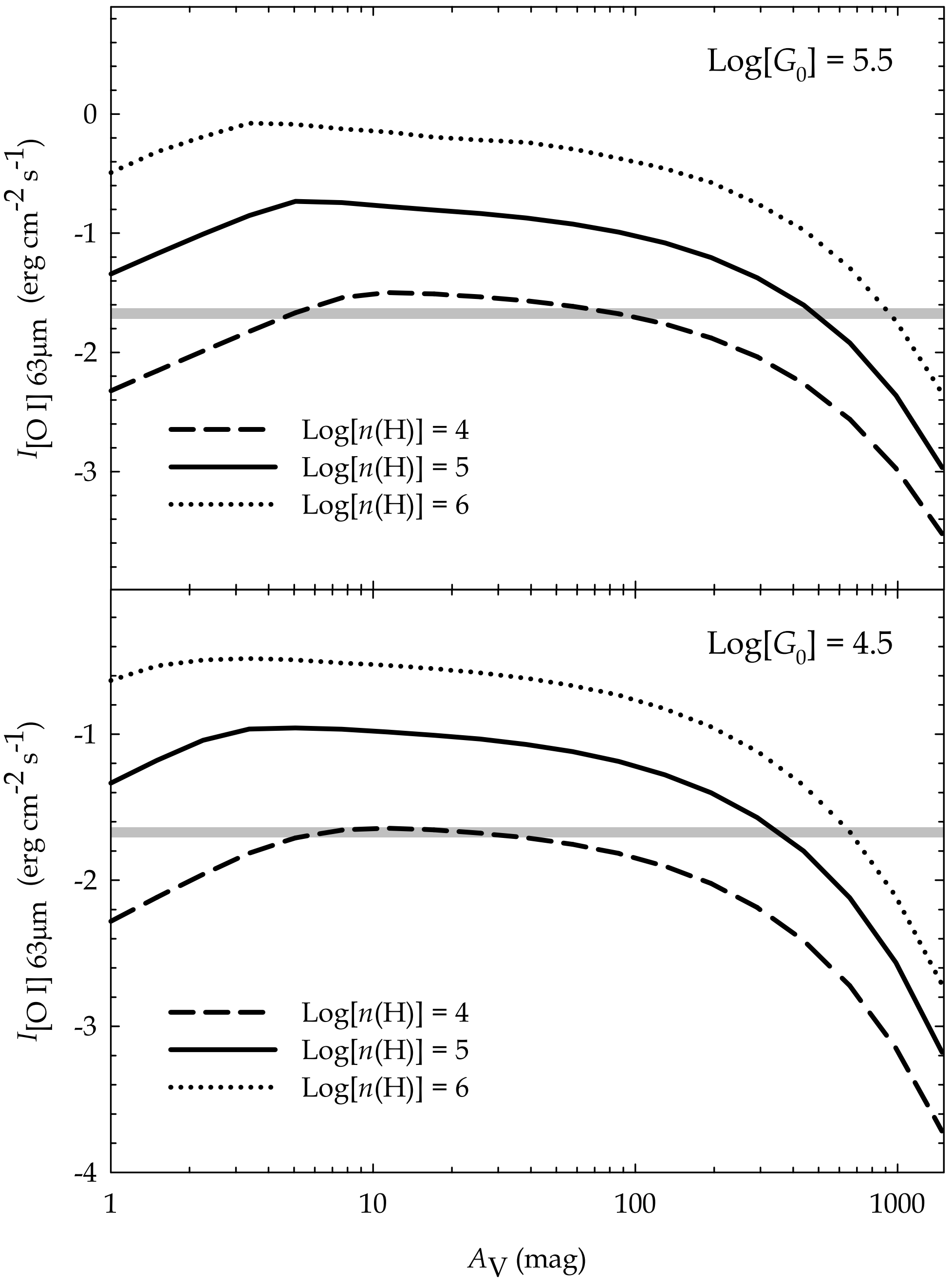

Figure 4 Emergent intensity of [O I] 63μm line for $n_H = 10^{4.0}$, $10^{5.0}$, and $10^{6.0}$ cm$^{-3}$ and $G_0 = 10^{4.5}$, $10^{5.5}$ as a function of $A_V$. The horizontal gray bar represents the observed 2σ range. The intensity increases with increasing $n_H$ or $G_0$, as expected. For larger $A_V$, the increased line optical depth reduces the observed intensity. For $A_V$ = 1477 mag, the 63μm optical depth is ~250.

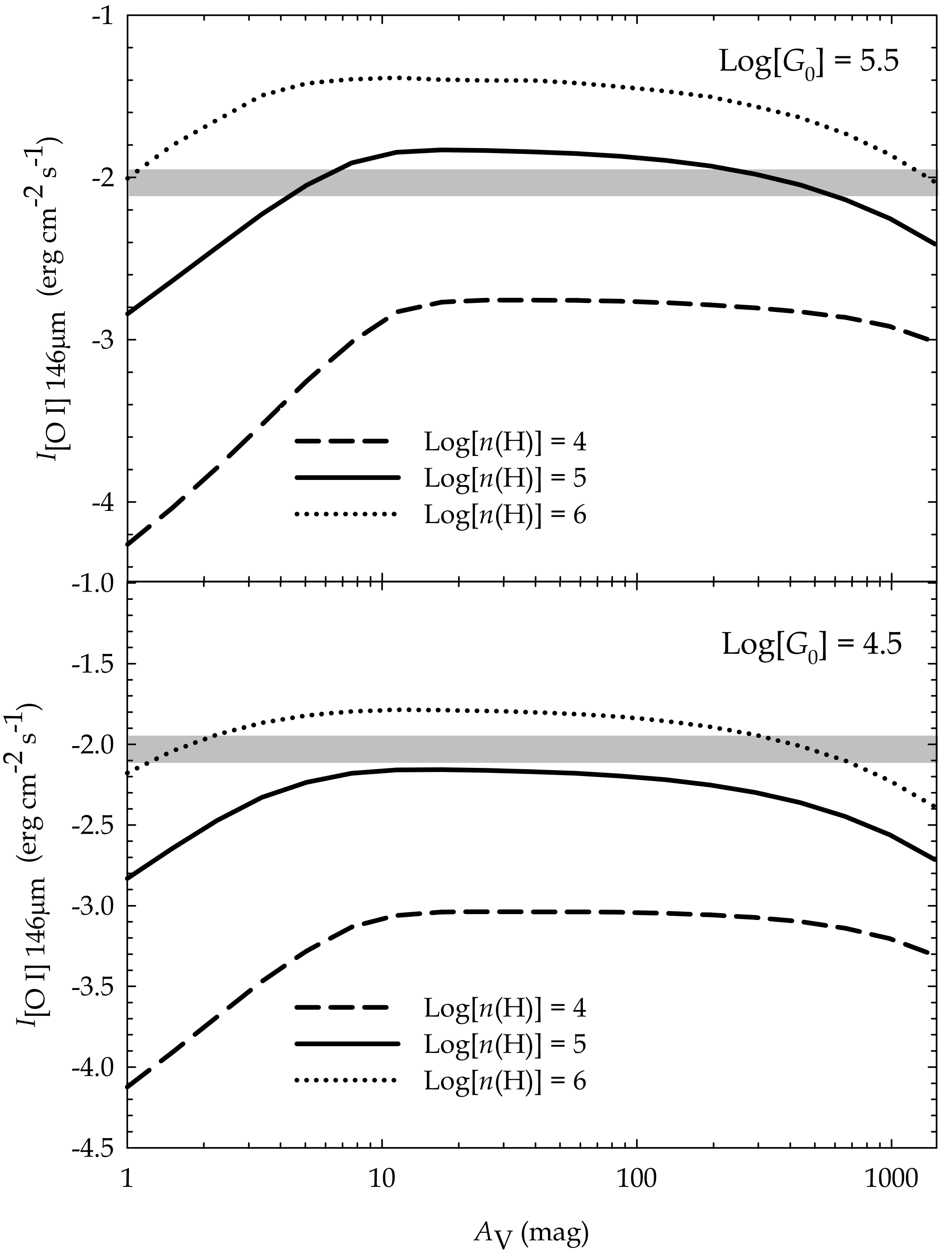

Figure 5 Emergent intensity of the [O I] 146μm line for $n_H = 10^{4.0}$, $10^{5.0}$, and $10^{6.0}$ cm$^{-3}$ and $G_0 = 10^{4.5}$, $10^{5.5}$ as a function of $A_V$. The horizontal gray bar represents the observed 2σ range. The intensity increases with increasing $n_H$ or $G_0$, as expected. For larger $A_V$, the increased line optical depth reduces the observed intensity, however line optical depth effects are much less important for this line than for 63μm emission. For $A_V$ = 1477 mag, the 146μm optical depth is ~2.5.

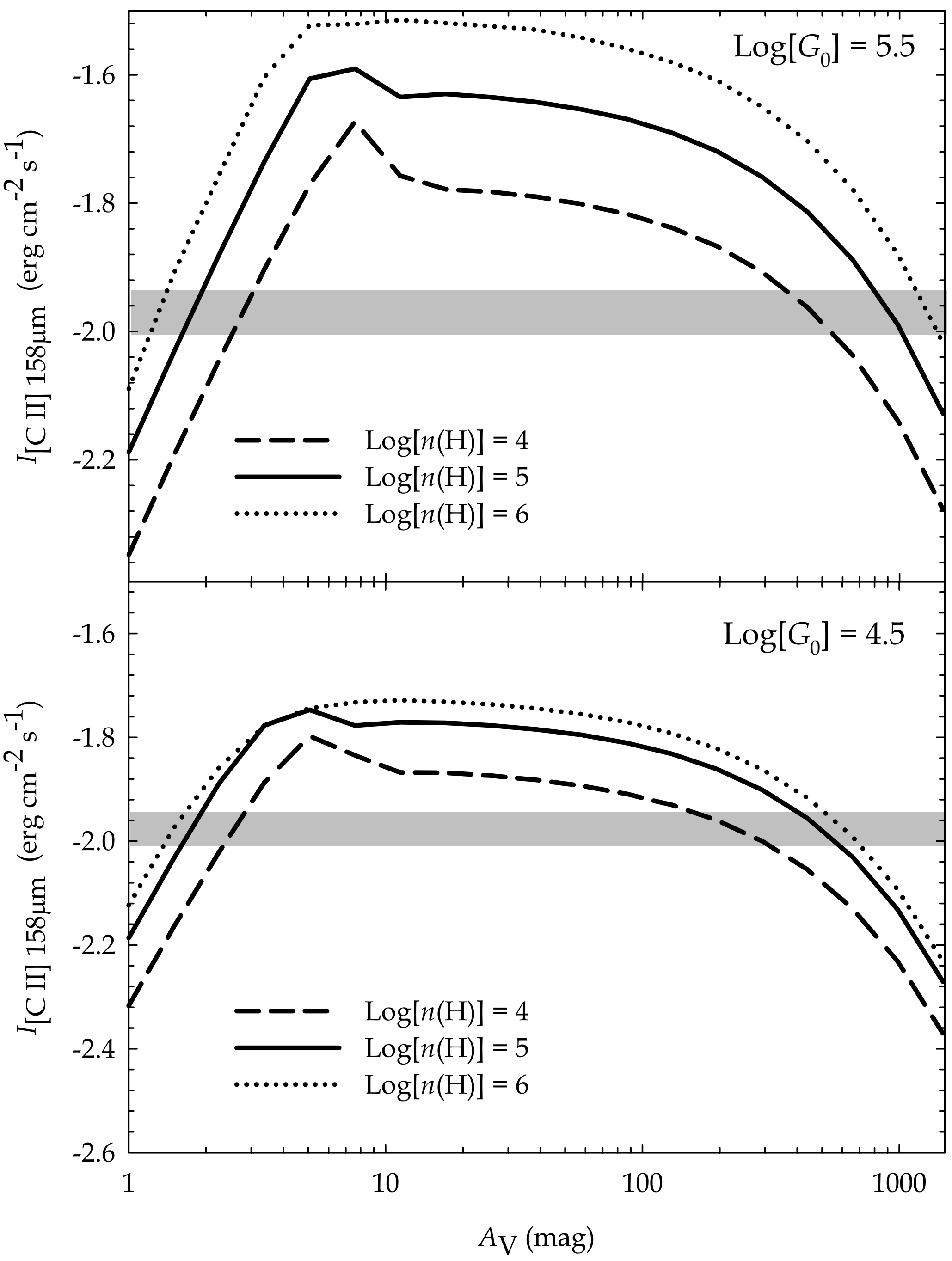

Figure 6 Intensity of [C II] 158μm line for $n_H$ = $10^{4.0}$, $10^{5.0}$, and $10^{6.0}$ $cm^{-3}$ and $G_0$ = $10^{4.5}$, $10^{5.5}$ as a function of $A_V$. The horizontal gray bar represents the observed 2σ range. The intensity increases with increasing $n_H$ or $G_0$, as expected. For larger $A_V$, the increased line optical depth reduces the observed intensity. This effect is much less than for 63μm emission, since most carbon is not in the form of $C^+$ but rather $C^0$ and CO. For $A_V$ = 1477 mag, the 158μm optical depth is ~1.6.

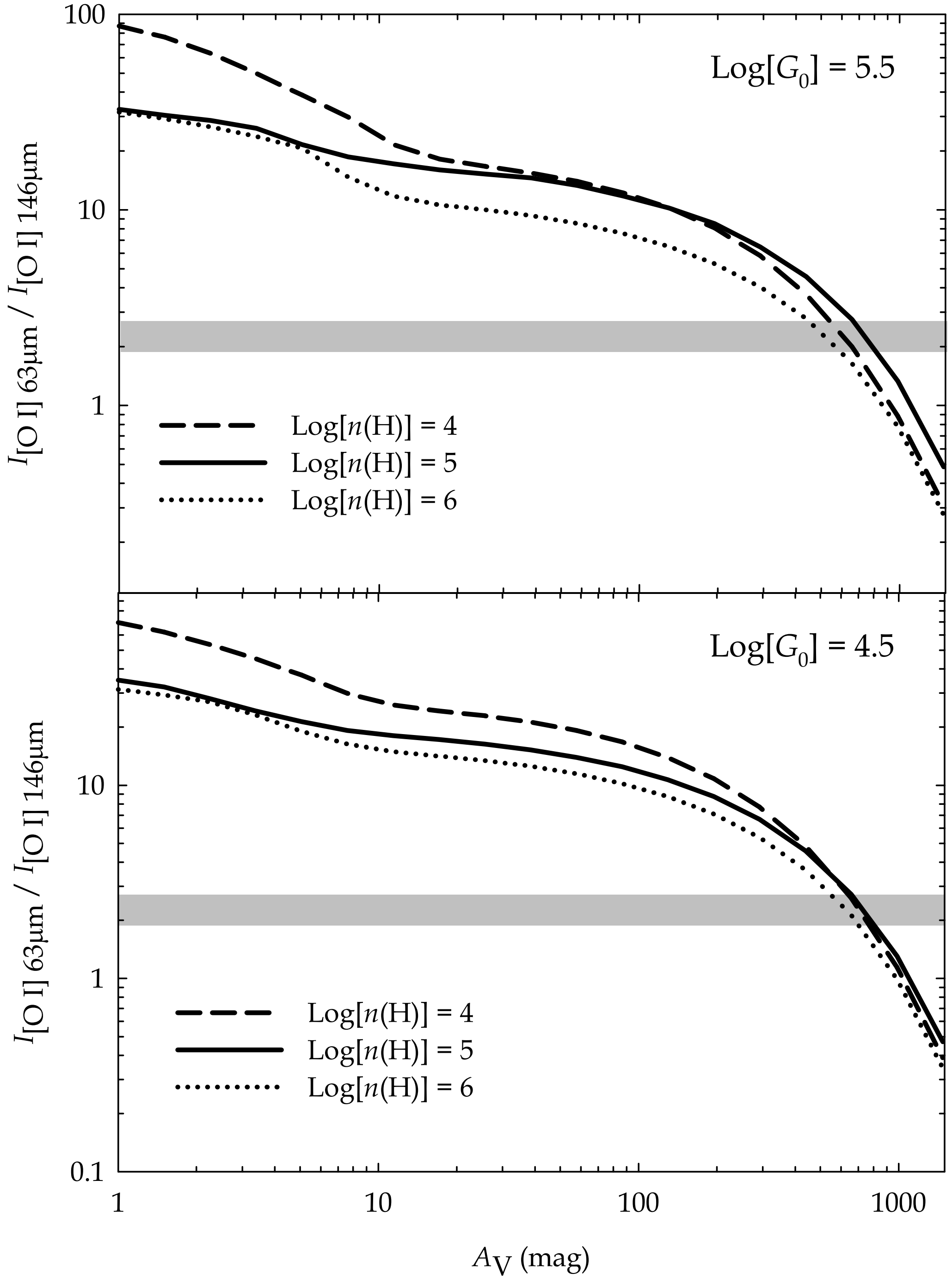

Figure 7 Ratio of the [O I] 63/146 line intensities for $n_H = 10^{4.0}$, $10^{5.0}$, and $10^{6.0}$ $cm^{-3}$ and $G_0 = 10^{4.7}$, $10^{5.7}$ as a function of $A_V$. The horizontal gray bar represents the observed 2σ range. To a good approximation, 63/146 is independent of $G_0$ for the range of $n_H$ and $G_0$ considered. As the amount of extinction increases, the $O^0$ column density increases. This leads to a larger [O I] 63μm optical depth and a reduction in 63/146.

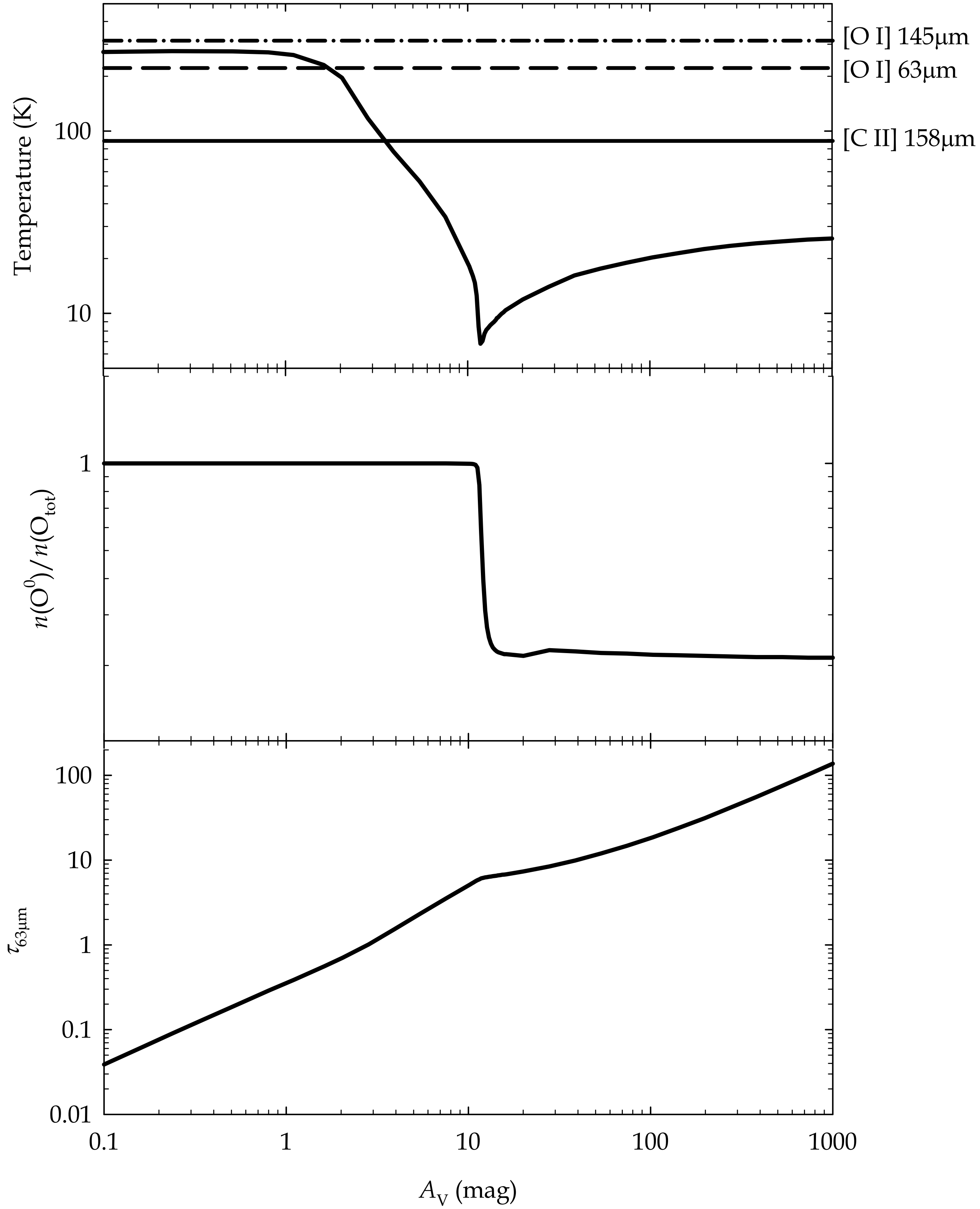

Figure 8 Temperature, $O^0$ fraction, and [O I] 63μm optical depth as a function of $A_V$ for the best-fit model for a constant density, closed geometry model given by Table 2. The horizontal lines on the temperature plot represents the temperature required to reach the upper level of the [C II] 158μm and [O I] 63μm, 146μm transition.